\documentclass[journal]{IEEEtran}
\ifCLASSINFOpdf
\else
\fi

\usepackage{siunitx}
\usepackage[pdftex]{graphicx}
\usepackage{amsmath}
\usepackage{shuffle}
\usepackage{stmaryrd}
\usepackage{tabu}
\usepackage[table]{xcolor}
\usepackage{amsthm}
\usepackage{import}
\usepackage{comment}

\usepackage{biblatex}
\begin{document}
%
\title{IQ Photonic Receiver for Coherent Imaging with a Scalable Aperture}
%
%
%

\author{Aroutin~Khachaturian,~\IEEEmembership{Member,~IEEE,}
        Reza~Fatemi,~\IEEEmembership{Member,~IEEE,}
        and~Ali~Hajimiri,~\IEEEmembership{Fellow,~IEEE}
\thanks{A. Khachaturian, Reza Fatemi, and Ali Hajimiri are with the Department
of Electrical Engineering, California Institute of Technology, Pasadena, 
CA, 91125 USA e-mail: akhachat@caltech.edu.}
\thanks{Manuscript received July xx, 2021; revised August xx, 2021.}}

%
%
\IEEEspecialpapernotice{(Invited Paper)}
\maketitle

\markboth{Open Journal of Solid-State Circuits Society,~Vol.~X, No.~Y, August~2021}%
{Shell \MakeLowercase{\textit{et al.}}: Bare Demo of IEEEtran.cls for IEEE Journals}
%



\maketitle

\begin{abstract}
Silicon photonics (SiP) integrated coherent image sensors offer higher sensitivity and improved range-resolution-product compared to direct detection image sensors such as CCD and CMOS devices. Previous generation of SiP coherent imagers suffer from relative optical phase fluctuations between the signal and reference paths, which results in random phase and amplitude fluctuations in the output signal. This limitation negatively impacts the SNR and signal acquisition times. Here we present a coherent imager system that suppresses the optical carrier signal and removes non-idealities from the relative optical path using a photonic in-phase (I) and quadrature (Q) receiver via a $90^\circ$ hybrid detector. Furthermore, we incorporate row-column read-out and row-column addressing schemes to address the electro-optical interconnect density challenge. Our novel row-column read-out architecture for the sensor array requires only $2N$ interconnects for $N^2$ sensors. An $8\times8$ IQ sensor array is presented as a proof-of-concept demonstration with $1.2\times 10^{-5}$ resolution over range accuracy. Free-space FMCW ranging with \SI{250}{\micro \meter} resolution at \SI{1}{\meter} distance has been demonstrated using this sensor array.
\end{abstract}


%
\IEEEpeerreviewmaketitle

%
%
%
%
\section{Introduction}

\IEEEPARstart{O}{ptical} imagers have a wide range of applications in microscopy, medical imaging, remote sensing, and 3D imaging (LIDAR) such as autonomous vehicles, robotics, and surface metrology. Traditional CCD and CMOS image sensors have performance limitations due to the their dark current noise and read noise (input-referred-noise) in low-light conditions. On the other hand, coherent heterodyne detection offers improved sensitivity due to the perceived gain of the reference path, which enables coherent imagers to operate close to the shot-noise limited regime. Furthermore, the advancement of integrated optical processing technologies, such as silicon photonics (SiP) platforms, permits compact and complex waveform processing. This allows coherent photonic imagers to benefit from signal enhancement techniques available in RF heterodyne receivers and radar systems, while having a smaller aperture and a superior spatial resolution offered by the photonics micrometer-scale wavelengths found in photonics compared to their RF counterparts. Furthermore, coherent receivers have been utilized for remote sensing, LiDAR \cite{LIDARev:2020, Gao:12}, and medical imaging applications (such as OCT \cite{Eggleston:18}) via loss, refractive index, or time-of-flight measurements.
\par
Coherent imagers operate either with flash illumination \cite{LidarSurvey:Flash, LincolnLabFlash:2016} or on a point-by-point basis using electrical \cite{AflatouniPrj:15, Wagner:19}, or mechanical steering \cite{Kirmani58, TUANTRANONT:01, Wang:19}. Point-by-point illumination improves the SNR and reduces the interference \cite{Wang:20}. As a result, point-by-point illumination has been the subject of much attention with the introduction of all-integrated optical phased arrays (OPA), which allow for solid-state beam steering \cite{Reza:JSSC, Watts:Nature, Miller:18}. On the receiver end, a lens \cite{Aflatouni:15,Behroozpour} or a solid-state beamforming receiver \cite{Fatemi:17} reconstructs the image of the target, and signal detection is achieved via heterodyne mixing. 

\par
\begin{figure}[!t]
	\includegraphics[width=1\linewidth]{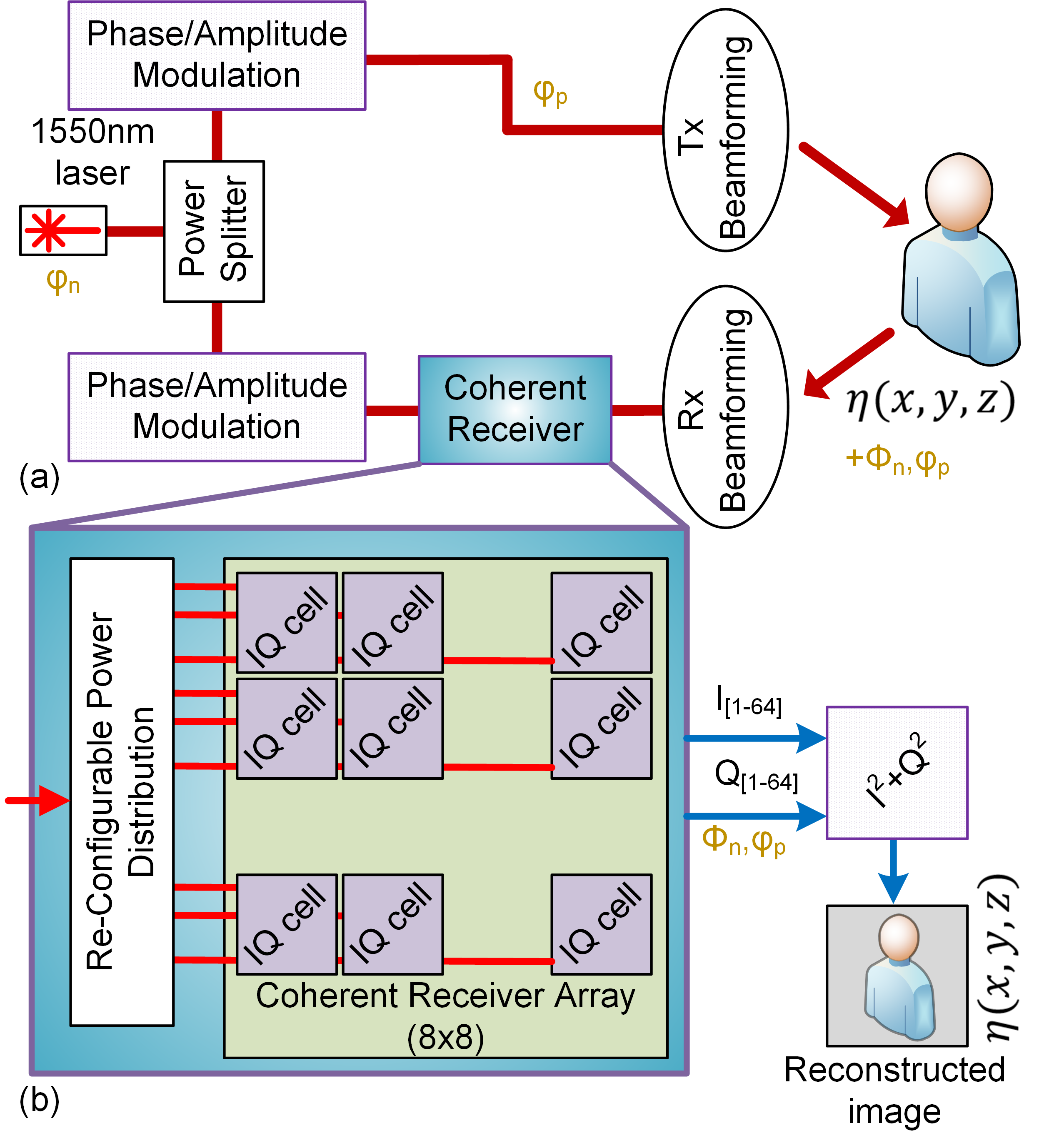}
	\caption{IQ coherent imager. (a) Typical coherent imaging scenario. (b) Proposed IQ imager with a scalable aperture.} 
	\label{Overview}
\end{figure}

Such an imaging system is shown in Fig. \ref{Overview}(a). The desired image information, whether it is the refractive index, the optical absorption of the material, or time-of-flight ($\eta(x,y,z)$) can be extracted after digitization in post-processing \cite{Aflatouni:15, Behroozpour, rogers2021universal}. There are two challenges with these architectures. First, the undesired relative optical phase fluctuations between the illumination and reference path due to the laser phase noise ($\phi_n$) or due to the thermal fluctuation of the transmit and the receive path, and target movement ($\phi_p$) translates into output signal phase and amplitude fluctuations that can degrade the system's performance. Typically, in conventional heterodyne mixers, this problem is addressed by increasing the signal acquisition time. This problem also occurs in RF coherent receivers and is addressed via IQ receivers. In optical receivers, whether they are used for communications \cite{Dong:14120hyb} or medical imaging applications (such as OCT \cite{SingleShotOCT}, LiDAR \cite{leeb1983}, and remote sensing \cite{Li:13}) structures such as $90^\circ$ hybrid combiners or $120^\circ$ hybrid combiners \cite{Kazovsky:87, leeb1983, Reiner:1991} are used to suppress the optical carrier signal. In this work, we have implemented a tunable $90^\circ$ hybrid architecture with balanced detectors. This integrated IQ receiver is analyzed in section \ref{IQtheory}.

\par

The other challenge with coherent imaging systems is scaling towards megapixel apertures, which is difficult due to the interconnect density limitations. For very large apertures, control and read-out signal routing to the inner elements of the array becomes a challenge. One method of addressing this problem is to use complex interconnect technologies such as through-silicon-vias (TSV) \cite{Kim:19}, or monolithic electronic-photonics platforms \cite{Chung:17,rogers2021universal}, which will add to the cost of the imaging system. We address the interconnect density challenge using a novel row-column read-out architecture based on diode-resistor logic \cite{wilkinson:1963} implemented in a photonic platform. This method reduces the read-out for $N^2$ pixels from $N^2$ interconnects to $2N$. A tunable reference distribution network further improves the dynamic range of the pixel interface via this row-column read-out architecture. This method significantly simplifies the analog amplification back-end and reduces the overall system power consumption. On the other hand, the interconnect density challenge for the control signals is addressed using a row-column drive architecture \cite{Reza:JSSC} which also reduces the number of interconnects for control from $N^2$ to $2N$ for $N^2$ pixels. The details of these designs are addressed in section \ref{IQScaling}.
\par

In this work, we demonstrate  a large-scale coherent imager architecture by implementing an $8\times8$-pixel IQ imager with a scalable aperture in a standard silicon photonics process. Our measurements demonstrate that this imager can measure frequency differences between the reference and illumination path with $1.2\times10^{-5}$ resolution over range accuracy within a \SI{1}{\milli \second} signal acquisition time. Our row-column read-out architecture exhibited better than \SI{-80}{\decibel} cross-talk below \SI{5}{\mega \hertz} frequency offset. We characterize the performance of this imager in a practical scenario by performing a free-space FMCW ranging measurement with this sensor, obtaining \SI{250}{\micro \meter} range resolution at \SI{1}{\meter} distance.

    \label{IQIntro}
    
\section{Photonics IQ Receiver}

\begin{figure}[!t]
	\includegraphics[width=1\linewidth]{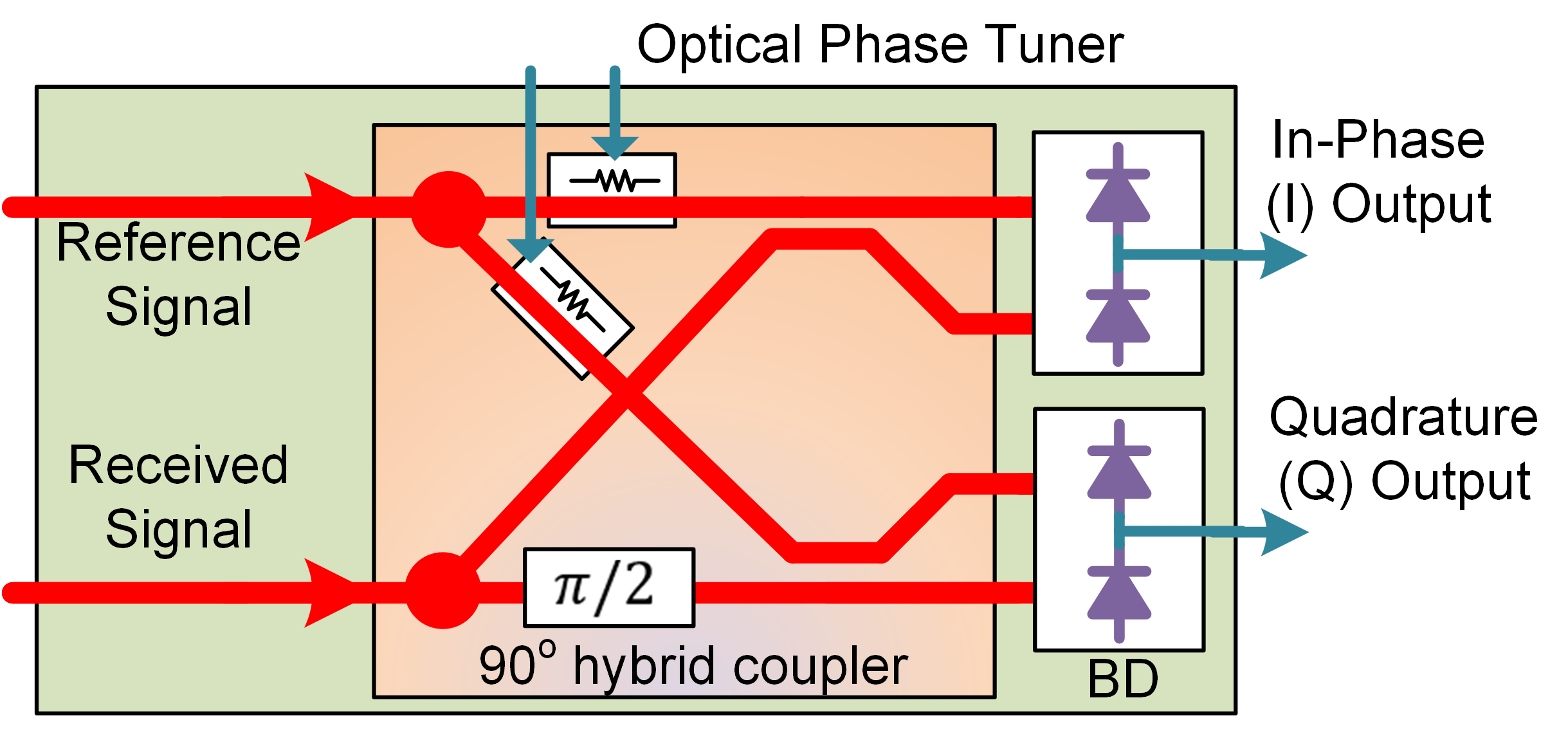}
	\caption{$90^\circ$ hybrid detector. A directional-coupler based $90^\circ$ hybrid ensures low-loss in-phase and quadrature signal generation. Balanced detectors suppress the common-mode signal.} 
	\label{NDHD}
\end{figure}

\begin{figure} [!b]
	\includegraphics[width=\linewidth]{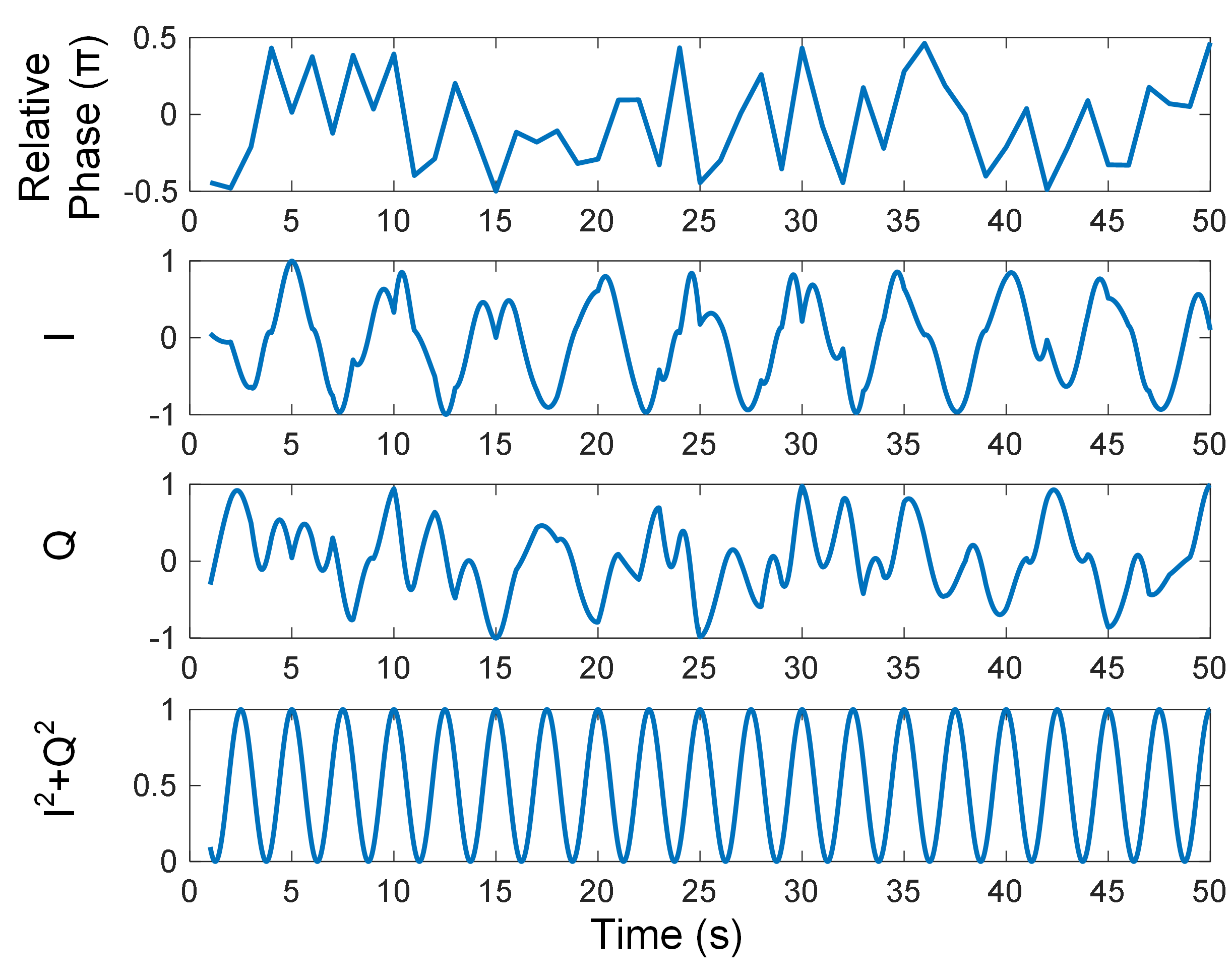}
	\caption{Simulation example of optical phase fluctuation suppression using IQ receiver.}
	\label{IQsim}
\end{figure}

As mentioned earlier, heterodyne detection in photonic coherent detection offers improved SNR by eliminating the contribution of read-noise and the detector's dark-current noise. The current signal at the output of the detector is the down-converted mixed component of the received and reference signals,  
\begin{equation}
    E_r(t)=\eta A_1(t)cos(\omega_{opt}(t)+B_1(t)+\phi)
    \label{E_rec} 
\end{equation}
\begin{equation}
    E_{LO}(t')=A_2(t')cos(\omega_{opt}(t')+B_2(t')+\phi')
    \label{E_LO} 
\end{equation}

where $A_i(t)$ and $B_i(t)$ are amplitude and phase modulation signals, respectively, and $\eta$ is the complex reflection coefficient of the target. Assuming a photodetector responsivity of $R$, the output current is given by:
\begin{equation}    \label{HeterodyneMix} 
    \begin{split}
    I(t) &= R\eta^2A_1^2(t)+RA_2^2(t') \\
         &+2RA_1(t)A_2(t')\eta cos(B_1(t)-B_2(t')+\Delta \phi),
    \end{split}
\end{equation}
where $\Delta \phi=\phi-\phi'$ is the contribution of the laser phase noise and receiver/illumination signal path phase mismatches. The desired target information is contained in the third term of Eq. \ref{HeterodyneMix}. The first two terms of this equation can be subtracted using balanced detectors. If phase modulators are used for time-of-flight measurements (such as FMCW) the mixed signal will simplify to
\begin{equation}    \label{Eq_PM} 
    I(t) = 2\eta R cos(B_1(t)-B_2(t')+\Delta \phi),
\end{equation}
where the path mismatch term $\Delta \phi$ will degrade the signal term $B_1(t)-B_2(t')$ as phase-noise. In this case, time-domain variations in $\Delta \phi$ will be indistinguishable from the signal term. In the case that amplitude modulators are used for time-of-flight measurements, the mixed-signal Eq. \ref{HeterodyneMix} will simplify to
\begin{equation}    \label{Eq_AM} 
    I(t) = 2\eta R A_1(t)A_2(t')cos(\Delta \phi),
\end{equation}
where the path mismatch term $\Delta \phi$ will degrade the mixed signal term $A_1(t)A_2(t')$ as an amplitude noise. In both these cases, it is desired to completely remove the optical path phase difference term, $\Delta \phi$, from the signal.
\par
Instead of standard heterodyne mixing, which is prone to fluctuations in the optical carrier phase signal, the output signal can be broken down to in-phase and quadrature components using the $90^\circ$ hybrid detector as shown in Fig. \ref{NDHD}. The reference signal and the received signal are split using a \SI{3}{\decibel} coupler and combined using directional couplers in two separate paths with a $90^\circ$ optical delay difference in the signal paths. The resulting signals are detected using balanced detectors to remove the common-mode terms. The two electrical outputs of the hybrid $90^\circ$ detectors are:
\begin{equation}
    I(t)= \eta RA_1(t)A_2(t')\eta cos(B_1(t)-B_2(t')+\Delta \phi)
    \label{Isig} 
\end{equation}
\begin{equation}
    Q(t)= \eta RA_1(t)A_2(t')\eta sin(B_1(t)-B_2(t')+\Delta \phi)
    \label{Qsig} 
\end{equation}
\par
Computing the sum of the square of these two signals, \ref{Isig} and \ref{Qsig}, removes the optical path phase mismatch term $\Delta \phi$ term resulting in 
\begin{equation}
    I^2(t)+Q^2(t)= \eta^2R^2A_1^2(t)A_2(t')^2,
    \label{I2Q2sig} 
\end{equation}
which completely removes the amplitude fluctuations for the case of amplitude modulated time-of-flight measurements in Eq. \ref{Eq_AM}. The proposed hybird $90^\circ$ coupler in this design incorporates directional couplers, which have negligible loss compared to MMI couplers used in \cite{Jeong:10}, which exhibit an additional \SI{0.5}{\decibel} loss. Furthermore, small thermal modulators (Fig. \ref{NDHD}) can be incorporated in the path of the reference signals to correct for deviations from the ideal $90^\circ$ path difference due to fabrication mismatch, operational wavelength, or temperature change. A simulation of the output I, Q, and $I^2+Q^2$ signal is shown in Fig. \ref{IQsim}.

    \label{IQtheory}

\section{Scalable Aperture}
To create a low-complexity expandable aperture, we utilized a diode-resistor, row-column access logic to reduce the number of required amplifiers for an $N\times N$ array of coherent receivers from $N^2$ to $N$. This reduces system complexity and electrical power consumption and increases the aperture's fill factor. The imager's coherent pixels are arranged in a rectangular grid, with the balanced detectors' bias signals connected to a common row node, while the balanced detector output signals (I and Q per pixel) are connected to two common column nodes via a series silicon diode (available in photonics processes). The relative DC voltages at the row bias nodes and the column read nodes determine which silicon diode is forward biased to allow the balanced detector's mixed signal to be connected to the amplifier. For a given measurement cycle, all but one of the rows are in reverse bias. This blocks the signal path from the output of the balanced detectors in the other rows to the amplifiers. 

\begin{figure}[!t]
	\includegraphics[width=1\linewidth]{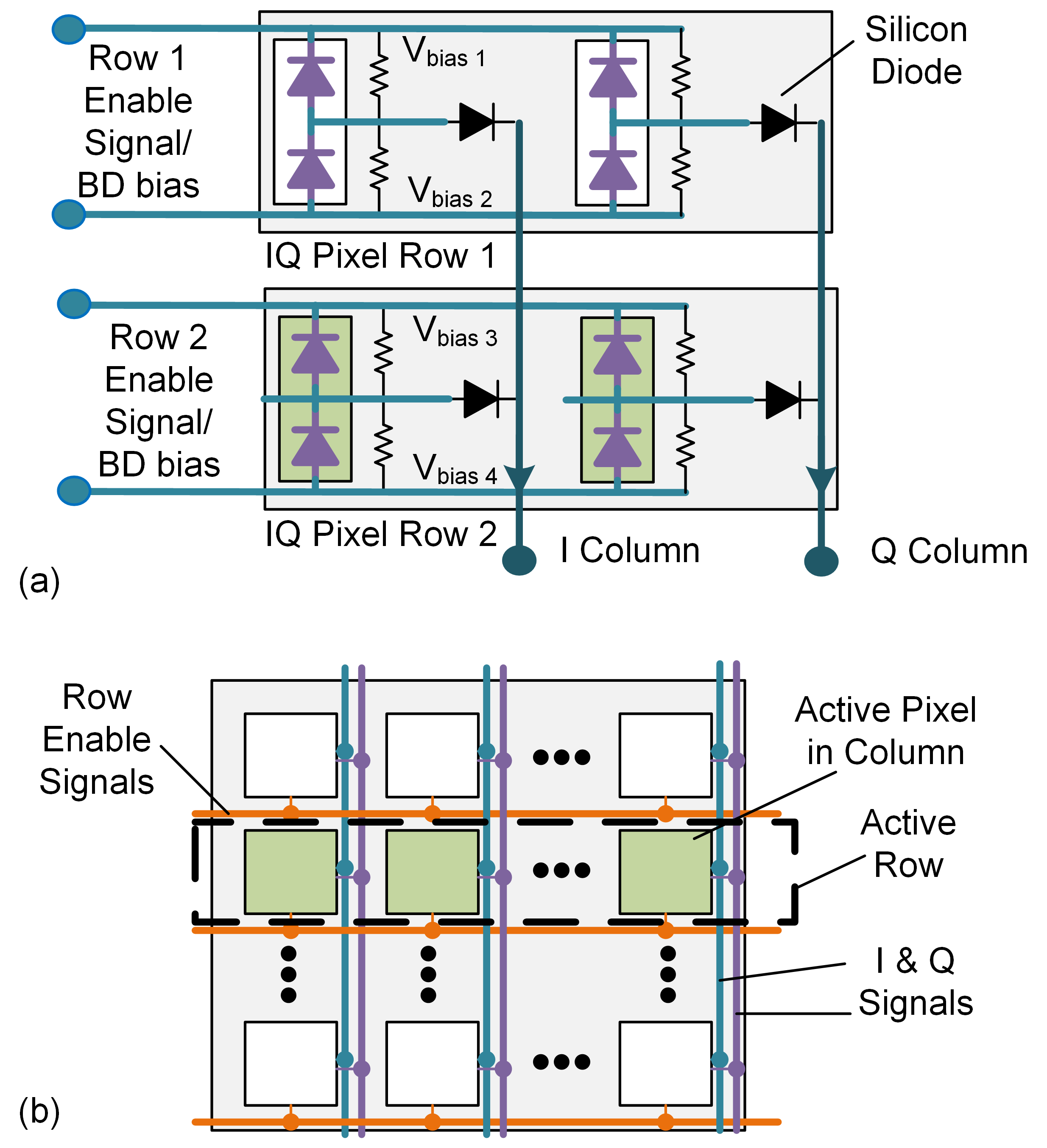}
	\caption{(a) Row-column read-out architecture. (b) Row-column read-out matrix for coherent imager.} 
	\label{IQscalability}
\end{figure}

\begin{figure}[!b]
	\includegraphics[width=1\linewidth]{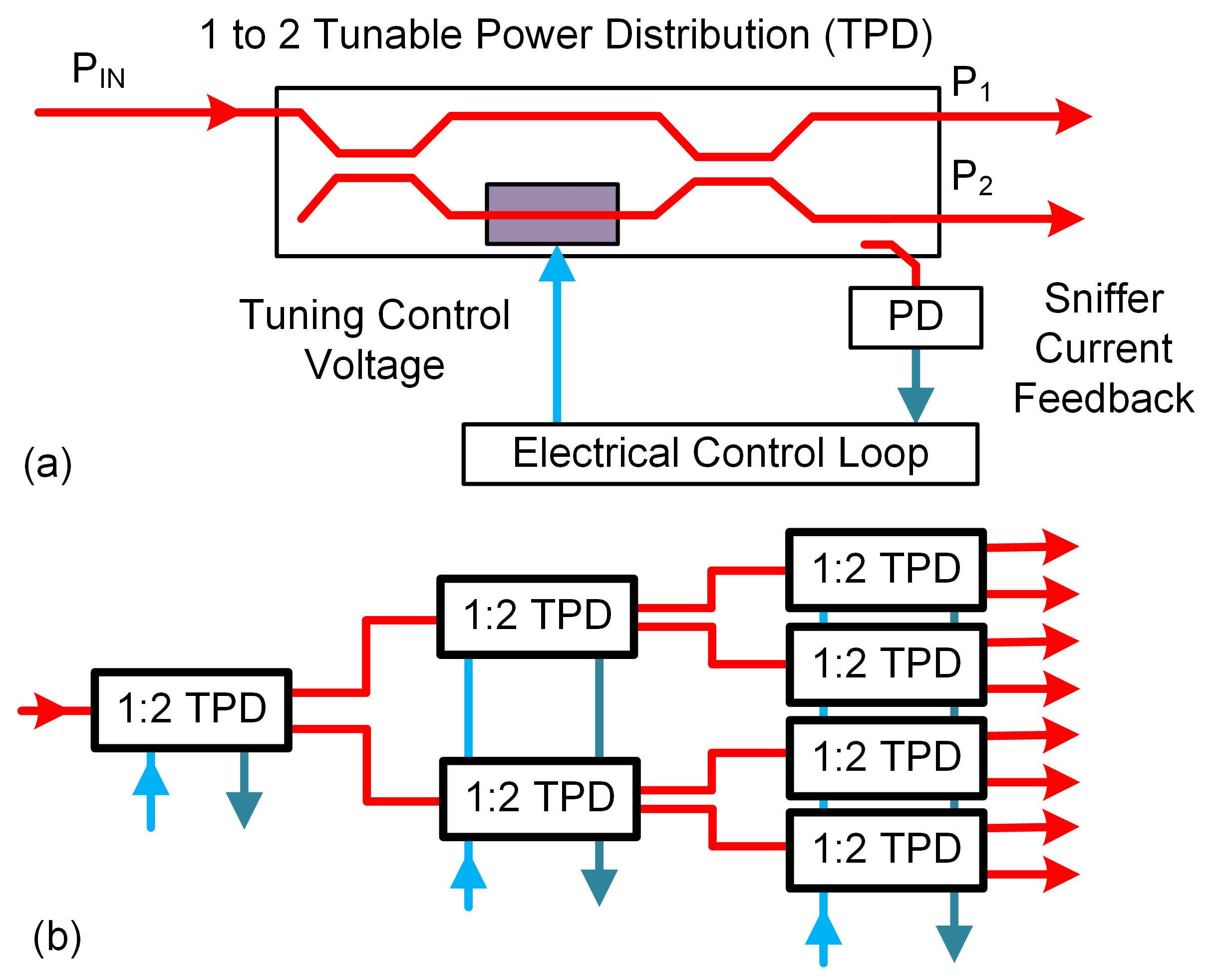}
	\caption{(a) Tunable 1:2 power distribution (b) 1:8 power distribution.} 
	\label{TPD}
\end{figure}

\par
The germanium photodetectors, that are disconnected because of the reversed-bias diode at a given read cycle, can get damaged due to charge build-up when there is no DC path for the generated charge in the diode to dissipate through \cite{qing:2015}. This problem is addressed by placing a parallel resistor with each photodiode that creates a current path for the turned-off rows.
\par

\begin{figure}[!t]
	\includegraphics[width=1\linewidth]{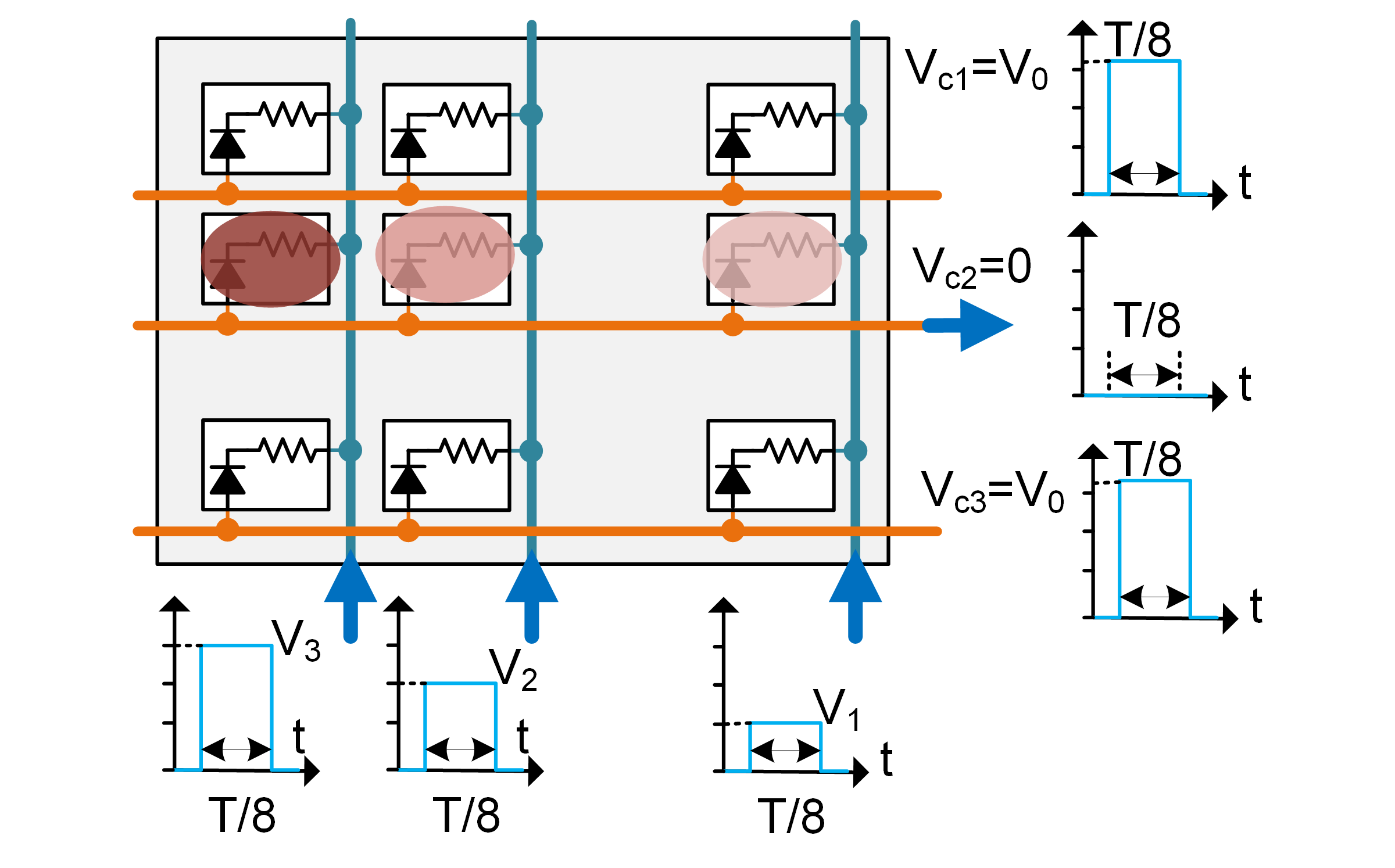}
	\caption{Row-column matrix addressing with PAM.} 
	\label{Fig_RCT}
\end{figure}

It can be observed that while the row-column read-out diode in reverse bias can provide electrical isolation between the rows, for very large arrays, the accumulated current leakage from the disabled rows can degrade the active pixel's dynamic range. To address this, we designed a tunable power distribution block (Fig. \ref{TPD}) that attenuates the reference power delivered to the disabled rows. This attenuates the downconverted mixed signal output from the turned-off rows. While this tunable power distribution can itself serve as a row selection mechanism, the combination of the diode-enabled rows with the tunable power distribution (TPD) ensures that the pixels will have a large dynamic range. The tunable coupler was implemented using a 1:2 (Fig. \ref{TPD}(a)) tunable Mach-Zhander interferometer. A small sniffer diode is included to monitor the power distribution ratio in a feedback loop. Cascading these TPDs creates a 1:8 tunable coupler network. 
\par
Another critical aspect for the scalability is row-column addressing of the thermal phase tuners (shown in Fig. \ref{NDHD}) in the 2D grid of IQ cells. To address this scalability challenge, we incorporated a row-column drive methodology utilizing the thermal memory of the phase shifters (typically in kHz range) as shown in Fig. \ref{Fig_RCT}. Each row of resistors in the thermo-optic phase shifters (TOPS) is activated by forward biasing the series diodes in that row for 1/8th of the cycle with pulsed-amplitude-modulation (PAM) drivers. All thermo-optic phase shifters receive $8P_{TOPS}$ the required power for 1/8 of the cycle. This enables independent programming of 64 phase shifters with only 16 electrical drivers. In addition, programming the rows at MHz frequencies ensures that the thermo-optic phase shifters receive a constant average power of $P_{TOPS}$.

    \label{IQScaling}

\section{Silicon Photonics Implementation}
The image sensor architecture was implemented in AMF's standard silicon photonics process. This proof-of-concept device contains an $8\times8$ coherent IQ pixel array as well as a 1:8 re-configurable amplitude distribution block with a calibration feedback (Fig. \ref{Overview}(b)). 
\par
\begin{figure}[!t]
	\includegraphics[width=1\linewidth]{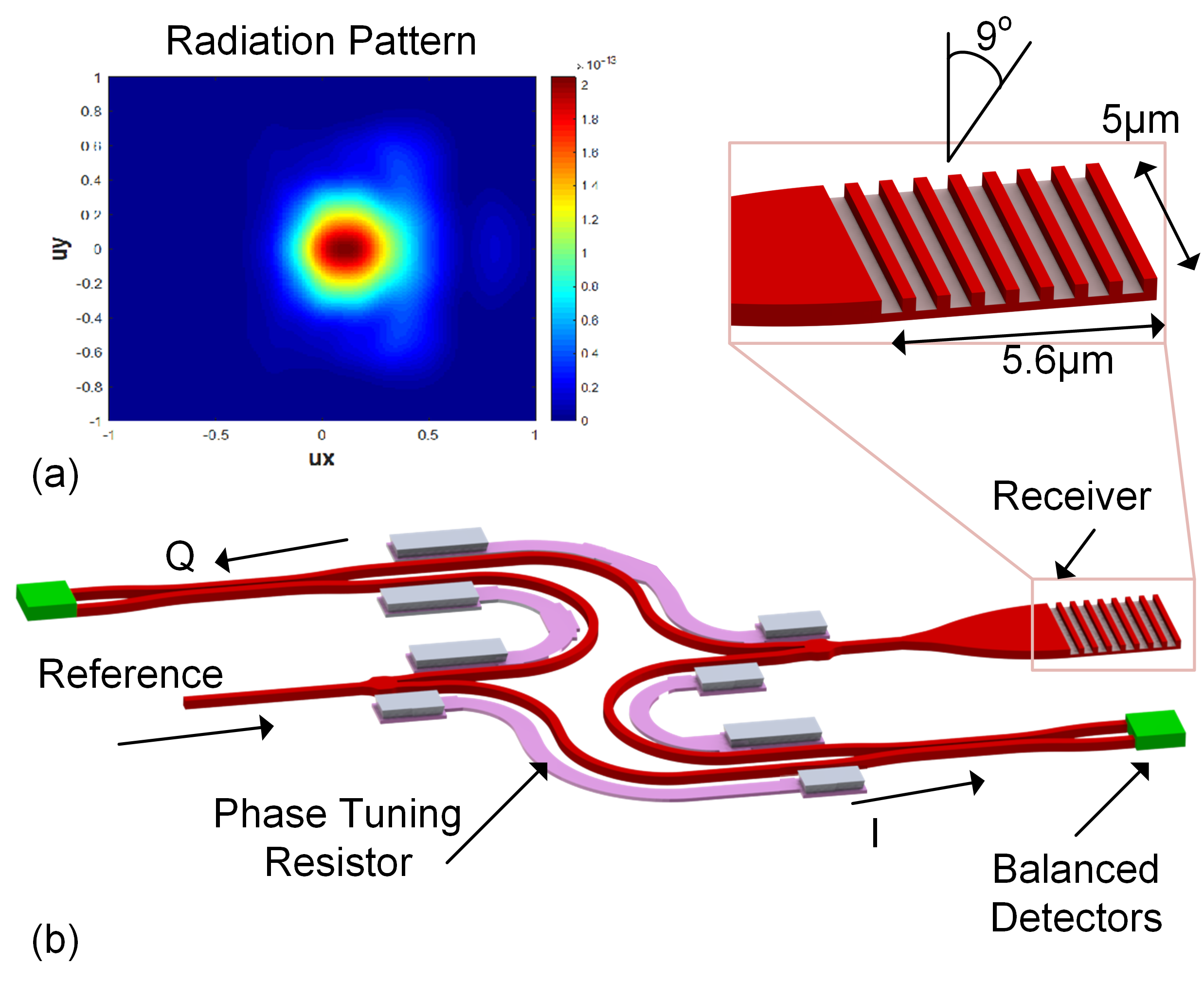}
	\caption{(a) Far-field pattern of the receiving element. (b) Hybrid $90^\circ$ detector implementation.} 
	\label{IQblocks}
\end{figure}

The receiver unit cell for IQ detection as shown in Fig. \ref{IQblocks}(a) splits the reference and received signal using \SI{3}{\decibel} y-junctions and mixes the two signals using directional couplers. The signal path length is adjusted by incorporating two symmetric $45^\circ$ waveguide length mismatches between the reference signal and the pixel for a total of $90^\circ$ path length different for I and Q generation. The outputs of each directional coupler are fed into a balanced detector to remove the common-mode signal at each output. For the row-column operation of this pixel, Fig \ref{IQscalability}, \SI{5}{\kilo \ohm} resistors were put in parallel with each photodiode to prevent damage via charge accumulation. A silicon diode was implemented using n-type and p-type dopings available for the silicon layer. Two small tuning resistors are included in the $90^\circ$ hybrid mixer to enable fine-tuning of the structure for changes in the operation wavelength and the fabrication mismatches. The two tuning resistors change the path length mismatch between the I and Q paths in the opposite direction. This enables IQ fine-tuning with very little power consumption. Each tuning resistor is broken down into two pieces for more uniform heating and is placed in series with a silicon diode for row-column addressing. As a result, the entire tuning resistor array of 128 thermo-optic phase shifters required only 24 electrical connections. The pixel pitch for this design was set to \SI{100}{\micro \meter}.
\par
The received signal was collected using a custom-designed and compact, $\SI{5.6}{\micro \meter}\times \SI{5}{\micro \meter}$, grating coupler as shown in Fig. \ref{IQblocks}(b). The simulated coupling efficiency of this grating coupler was \SI{-3}{\decibel} with an optimum angle of $9^\circ$ and a $10^\circ$ field-of-view.
\par
The tunable amplitude couplers were implemented using length-matched spiral thermo-optic modulators, similar to \cite{Reza:JSSC}, as shown in Fig. \ref{TPD}(a). Path-length matching reduces the thermal cross-talk via the substrate between tunable amplitude couplers. For this particular design, three stages of 1:2 tunable couplers enabled full amplitude tunability from the input to the eight outputs. This cascaded structure delivers a constant reference power to the receiver array and is more power efficient than dedicated amplitude modulators per row. The power received by each row is divided between the columns using a series of \SI{3}{\decibel} y-junction couplers. To calibrate the power requirements of the thermal modulators and correct for fabrication-related mismatches, a series of $1\%$ sniffer photodiodes are placed at one of the outputs of each tunable coupler. This results in eight sniffer detectors for this array. However, since calibration is required only once per chip, the sniffer diodes at each stage were connected in parallel as shown in Fig. \ref{TPD}(b), resulting in a total of three sniffer photodiode current outputs. Since the input power, $P_{in}$, is known, the output current of the first stage can be calibrated with the sniffer output of the first stage. Afterward, the entire power can be diverted to the top tunable coupler by adjusting the first tunable coupler, and the output of the second sniffer current can be used to calibrate the top tunable coupler. This process can be iterated for the remaining tunable couplers. This parallel connection of the sniffer detectors reduces the number of current outputs for calibration for a $1:2^N$ splitter from $2^N-1$ to $N$, which helps with the scalability of this architecture.

    \label{IQchip}

\section{Measurement}

\begin{figure}[!t]
	\includegraphics[width=1\linewidth]{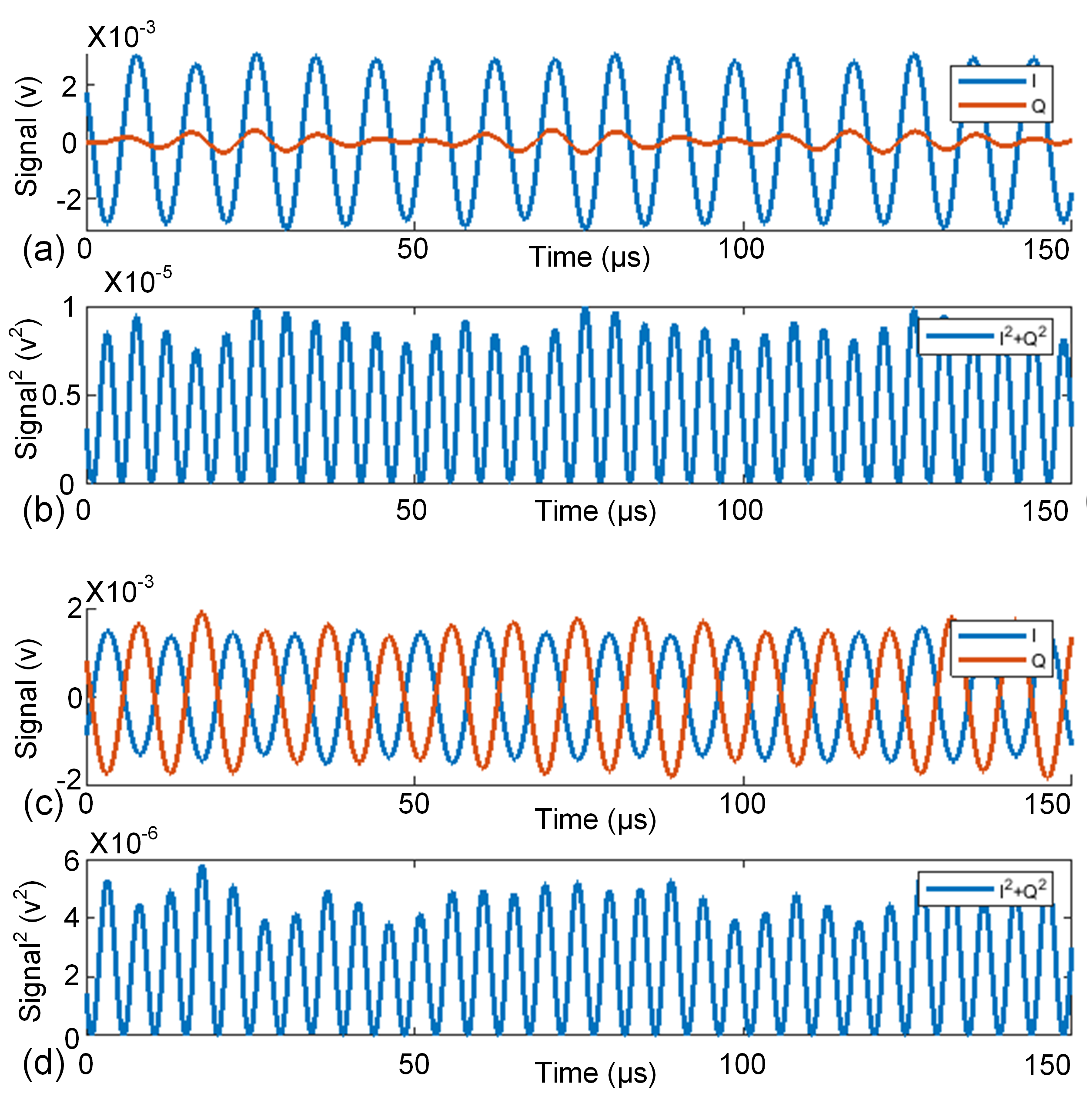}
	\caption{Samples of the measured IQ waveforms.} 
	\label{IQwaveform}
\end{figure}

\begin{figure}[!t]
	\includegraphics[width=1\linewidth]{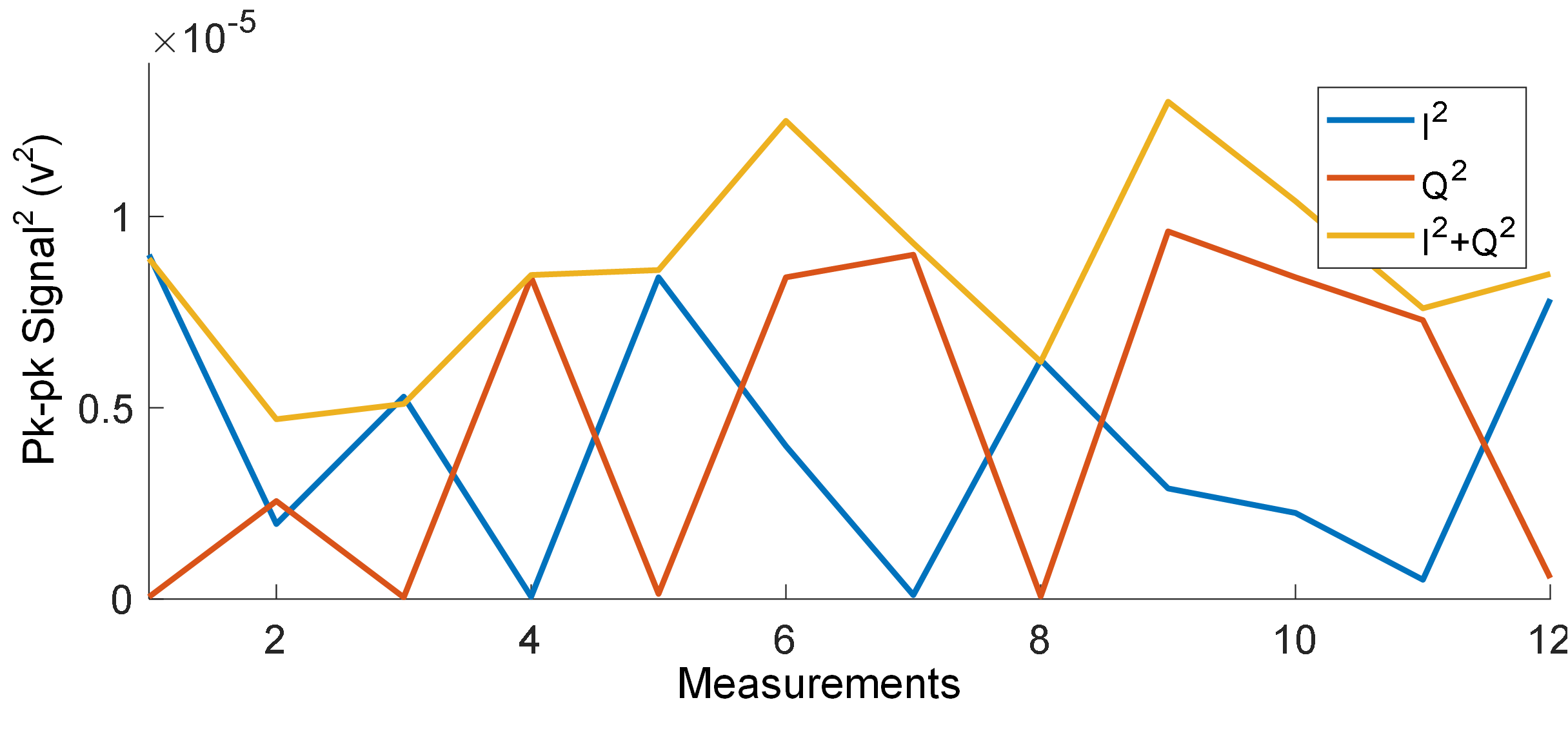}
	\caption{Peak-to-peak value of $I^2$, $Q^2$, and $I^2+Q^2$ signals over several measurements.} 
	\label{IQSNR}
\end{figure}

\begin{figure}[!t]
	\includegraphics[width=1\linewidth]{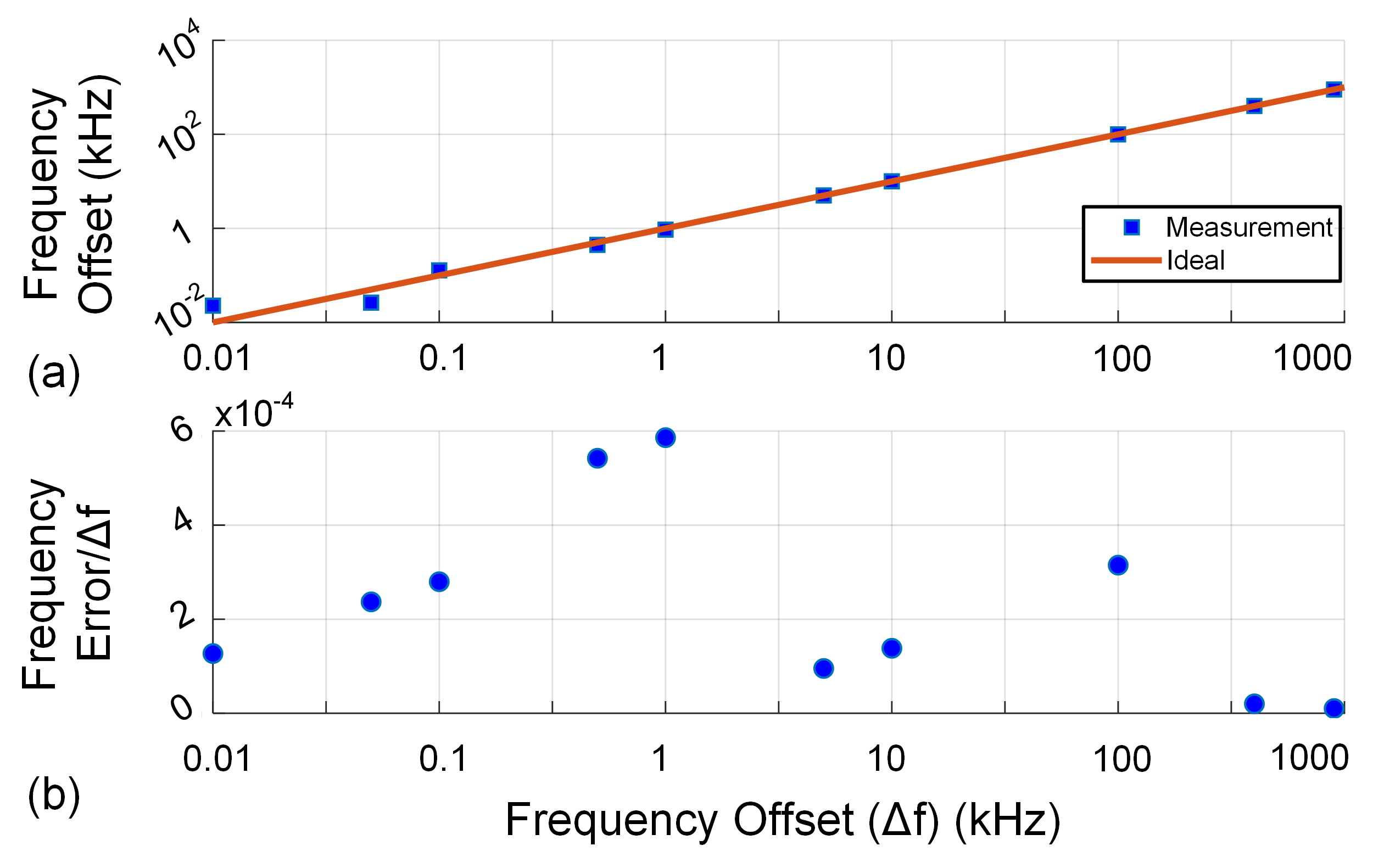}
	\caption{Frequency measurement accuracy. (a) Measured frequency vs pre-set frequency. (b) Ratio of the measured frequency error and the pre-set frequency.} 
	\label{IQres}
\end{figure}

For IQ imager characterization, a single pixel of the IQ imager was illuminated with fiber. A second fiber couples the reference light into the chip. The two signals were modulated externally at \SI{5}{\mega \hertz} and \SI{5.1}{\mega \hertz} with lithium-niobate amplitude modulators and amplified before coupling to the chip using erbium-doped fiber amplifiers (EDFAs). The row-column read-out voltages were adjusted such that only the particular pixel's row was active. The output mixed-signal current at \SI{100}{\kilo \hertz} modulation was amplified with an external transimpedance amplifier with \SI{5}{\kilo \ohm} gain. The resulting I and Q signals were digitized and filtered digitally with \SI{100}{\kilo \hertz} band-pass filters. Two examples of these waveforms are shown in Fig. \ref{IQwaveform}(a),(c). Due to the random phase fluctuations, there is a high chance that either of the I and Q signals will be attenuated (Fig. \ref{IQwaveform}(a)). Computing the sum-squared term in Eq. \ref{I2Q2sig} showed that while individual waveforms can exhibit amplitude fluctuations as a result of the relative optical phase fluctuations, the resulting sum-square term (Fig. \ref{IQwaveform}(b),(d)) exhibits little amplitude fluctuations and can suppress the effects of the optical carrier's phase. The experiment was repeated several times with different modulation frequencies to demonstrate this effect more clearly. The voltage swing of the $I$, $Q$, and $I^2+Q^2$ signals across these measurements are shown in Fig. \ref{IQSNR}. Individual I or Q signals might degrade by more than an order of magnitude; however, the sum squared term fluctuates around \SI{4}{\decibel} within our measurements. The results in Fig. \ref{IQSNR} further suggest that the peak-to-peak voltage swing on the sum-squared term is always larger than or equal to the peak-to-peak voltage swing of both channel. 
\par
Subsequently, we characterized the IQ imager's frequency resolving accuracy. Starting at a fixed frequency offset of $\Delta f = \SI{100}{\kilo \hertz}$ (modulating the two paths at \SI{5}{\mega \hertz} and \SI{5.1}{\mega \hertz}), we increased the frequency difference up to $\Delta f = \SI{1}{\mega \hertz}$ and measured the frequency difference from the optical data and compared it to the pre-set values from the arbitrary waveform generators. The result of this measurement is shown in Fig. \ref{IQres}(a). The I and Q signals were captured with \SI{1}{\milli \second} signal acquisition time and digitized. After filtering the signals with a \SI{100}{\kilo \hertz} digital filter, $I^2+Q^2$ term was calculated, and the mixed frequency was extracted from the data by measuring the zero-crossings of the mixed signal. For a frequency difference of \SI{1}{\mega \hertz}, which corresponds to 1000 cycles, the error in frequency measurement was \SI{12}{\hertz}. This error corresponds to a frequency measurement accuracy of $1.2\times10^{-5}$. The plot of the frequency measurement error normalized to the pre-set frequency is shown in Fig. \ref{IQres}(b). 

\begin{figure}[!t]
	\includegraphics[width=1\linewidth]{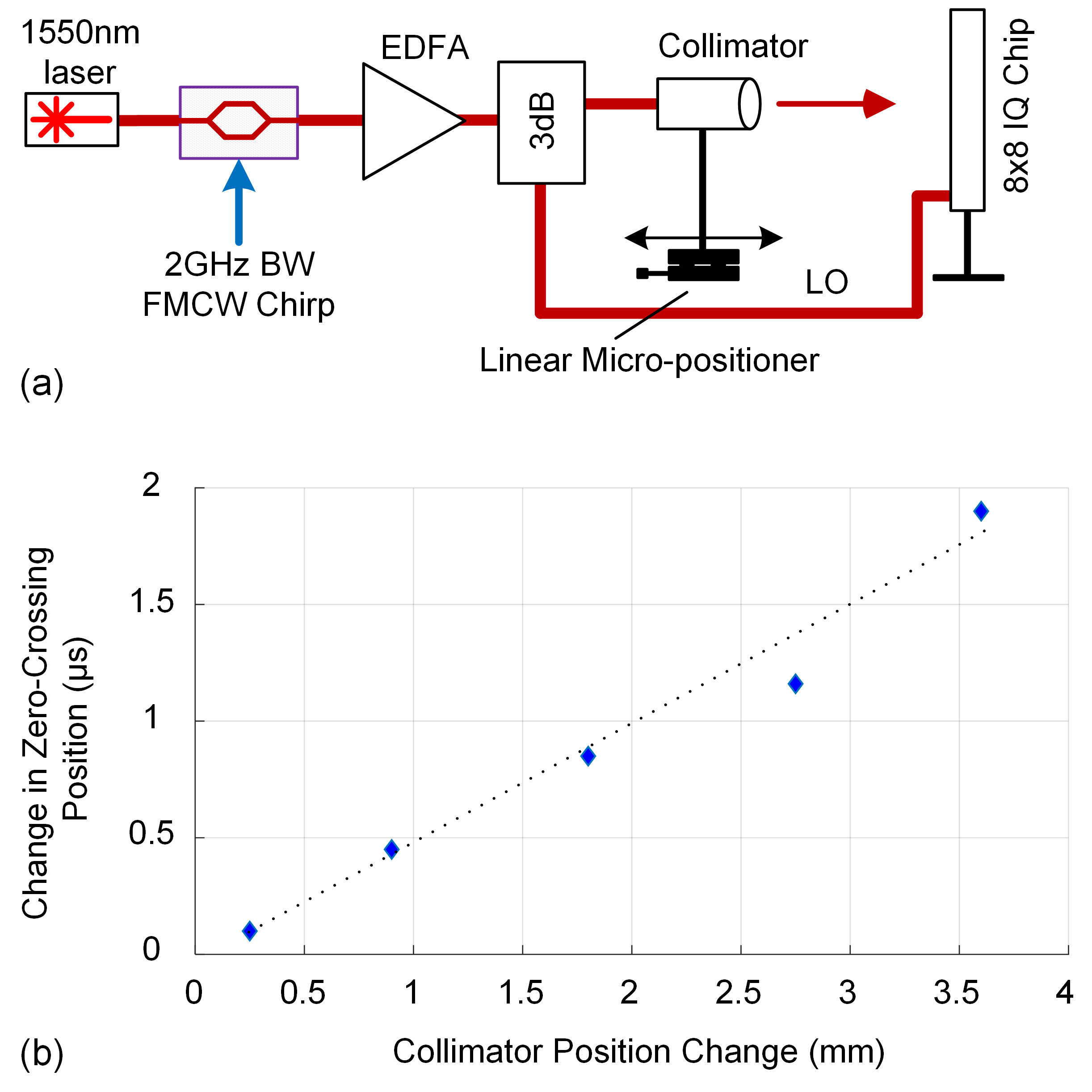}
	\caption{(a) IQ imager range measurement setup. (b) LIDAR range measurement.} 
	\label{LIDAR}
\end{figure}

\par
Afterward, to characterize the row-column read-out isolation, a pixel's mixed-tone power was measured when that row was enabled versus when it was disabled. At the system's default operation point with mixed frequency tone in \SI{100}{\kilo \hertz} to \SI{1}{\mega \hertz} range, no signal was observable when the row was disabled. At an increased offset frequency of \SI{5}{\mega \hertz} (from modulation frequency of \SI{6}{\mega \hertz} and \SI{11}{\mega \hertz}), we were able to measure \SI{-80}{\decibel} cross-talk from the disabled rows of the read-out circuit. This validates the high dynamic range of the proposed row-column read-out architecture.
\par
Finally, the IQ imager was characterized for LiDAR imaging applications as shown in Fig. \ref{LIDAR}(a). The image sensor was packaged and mounted on a stationary post. The pixels were illuminated using a collimator mounted on another post with linear movement controlled using a linear micro-positioner stage. The laser signal was modulated with an FMCW signal with \SI{2}{\giga \hertz} bandwidth and \SI{1}{\milli \second} chirp repetition-rate for \SI{2}{\tera \hertz \per \second} chirp rate. The modulated signal was amplified with an EDFA and split with a \SI{3}{\decibel} fiber splitter as illumination and reference signals. The illumination collimator was placed at a distance of \SI{1}{\meter} from the coherent imager, and its location was varied by several millimeters via the micro-positioner. The mixed down-converted signal was amplified and digitized. By measuring the zero-crossing point of the captured signal, we estimated the frequency of the mixed component and converted it to distance as shown in Fig. \ref{LIDAR}(b). We measured \SI{250}{\micro \meter} change in the distance at the setup's distance of \SI{1}{\meter}. This resolution and range ratio corresponds to a $2.5\times10^{-4}$ resolution over range accuracy for the FMCW measurement.
\par
\begin{figure}[!t]
	\includegraphics[width=1\linewidth]{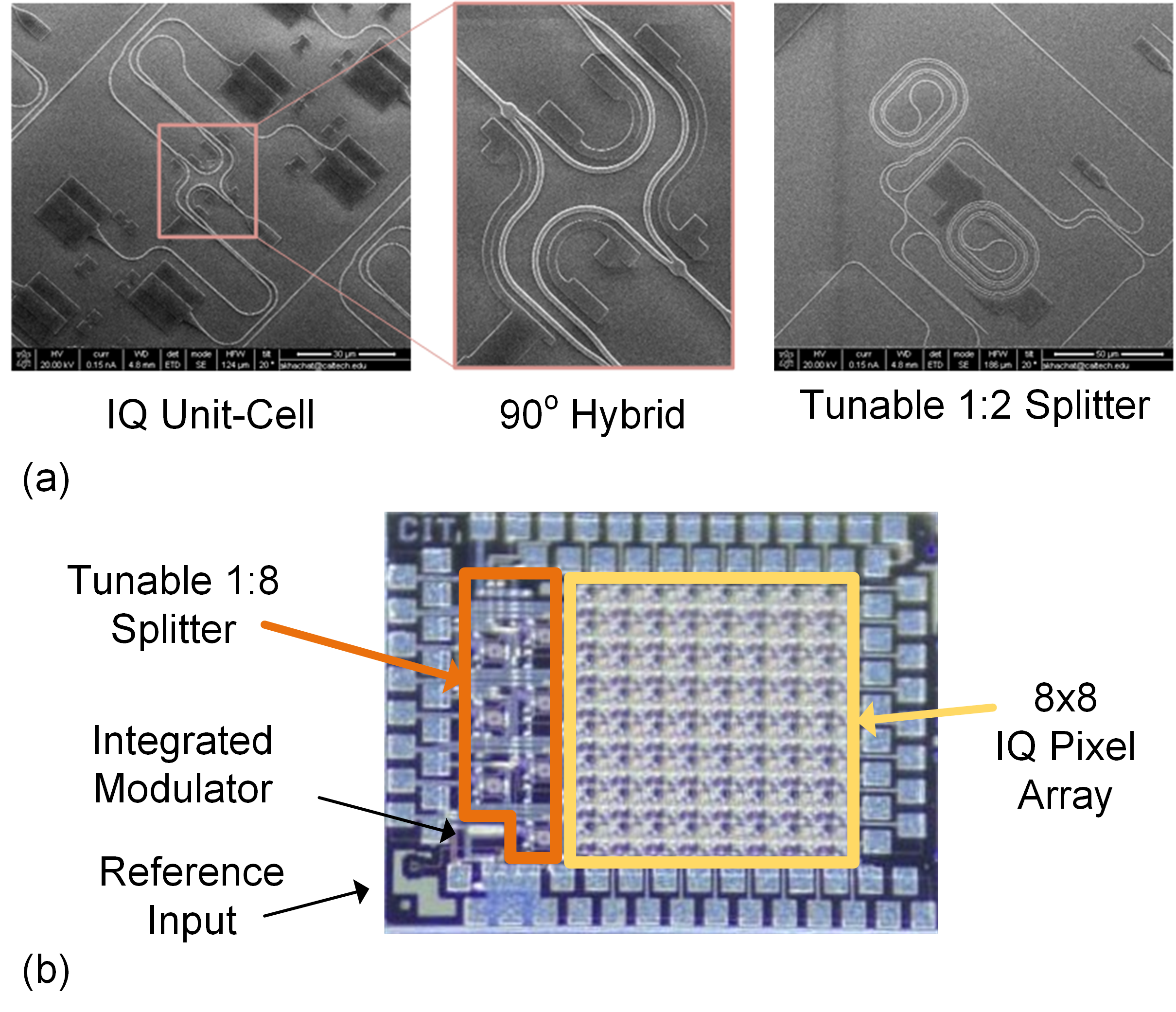}
	\caption{(a) SEM images of IQ building blocks. (b) IQ chip die photo.} 
	\label{Chip-img}
\end{figure}

SEM images of the coherent imager's IQ unit cell and tunable amplitude coupler are shown in Fig. \ref{Chip-img}(a). The die photo of the 8x8 coherent imager system is shown in Fig. \ref{Chip-img}(b).

    \label{IQmeasurements}
    
\section{Conclusion}

In this work, we demonstrated an integrated IQ coherent receiver with an expandable aperture. This receiver can suppress the optical carrier signal phase fluctuations and improves the system SNR. Furthermore, we demonstrated a novel row-column read-out architecture in a standard silicon photonics process which can  reduce the required electrical amplification and control system as well as reduce the electronic/photonic interconnect density requirements. Finally, we demonstrated LiDAR imaging, in a free-space measurement stage with \SI{250}{\micro \meter} resolution at \SI{1}{\meter} distance.

\section{Acknowledgement}
The authors would like to acknowledge Behrooz Abiri, Parham Porsandeh Khial, and Samir V. Nooshabadi for their valuable input and discussions.

\ifCLASSOPTIONcaptionsoff
  \newpage
\fi

\printbibliography

@INPROCEEDINGS{Behroozpour,
  author={B. {Behroozpour} and P. A. M. {Sandborn} and N. {Quack} and T. J. {Seok} and Y. {Matsui} and M. C. {Wu} and B. E. {Boser}},
  booktitle={2016 IEEE International Solid-State Circuits Conference (ISSCC)}, 
  title={11.8 Chip-scale electro-optical 3D FMCW lidar with 8um ranging precision}, 
  year={2016},
  volume={},
  number={},
  pages={214-216},
  doi={10.1109/ISSCC.2016.7417983}}

@article{Gao:12,
author = {S. Gao and R. Hui},
journal = {Opt. Lett.},
number = {11},
pages = {2022--2024},
publisher = {OSA},
title = {Frequency-modulated continuous-wave lidar using I/Q modulator for simplified heterodyne detection},
volume = {37},
month = {6},
year = {2012},
url = {http://ol.osa.org/abstract.cfm?URI=ol-37-11-2022},
doi = {10.1364/OL.37.002022}}

@inproceedings{Eggleston:18,
author = {Michael S. Eggleston and Flavio Pardo and Cristian Bolle and Bob Farah and Nicolas Fontaine and Hugo Safar and Mark Cappuzzo and C. Pollock and David J. Bishop and Mark P. Earnshaw},
booktitle = {Conference on Lasers and Electro-Optics},
journal = {Conference on Lasers and Electro-Optics},
pages = {JTh5C.8},
publisher = {Optical Society of America},
title = {90dB Sensitivity in a Chip-Scale Swept-Source Optical Coherence Tomography System},
year = {2018},
url = {http://www.osapublishing.org/abstract.cfm?URI=CLEO_AT-2018-JTh5C.8},
doi = {10.1364/CLEO_AT.2018.JTh5C.8}
}

@inbook{LidarSurvey:Flash,
author = {John A. Christian and Scott Cryan},
title = {A Survey of LIDAR Technology and its Use in Spacecraft Relative Navigation},
booktitle = {AIAA Guidance, Navigation, and Control (GNC) Conference},
chapter = {},
pages = {},
doi = {10.2514/6.2013-4641},
URL = {https://arc.aiaa.org/doi/abs/10.2514/6.2013-4641},
eprint = {https://arc.aiaa.org/doi/pdf/10.2514/6.2013-4641}
}

@article{Aflatouni:15,
author = {F. {Aflatouni} and B. {Abiri} and A. {Rekhi} and A. {Hajimiri}},
journal = {Opt. Express},
number = {4},
pages = {5117--5125},
publisher = {OSA},
title = {Nanophotonic coherent imager},
volume = {23},
month = {2},
year = {2015},
url = {http://www.opticsexpress.org/abstract.cfm?URI=oe-23-4-5117},
doi = {10.1364/OE.23.005117}}

@Article{Wang:20,
AUTHOR = {Wang, Dingkang and Watkins, Connor and Xie, Huikai},
TITLE = {MEMS Mirrors for LiDAR: A Review},
JOURNAL = {Micromachines},
VOLUME = {11},
YEAR = {2020},
NUMBER = {5},
ARTICLE-NUMBER = {456},
URL = {https://www.mdpi.com/2072-666X/11/5/456},
PubMedID = {32349453},
ISSN = {2072-666X},
DOI = {10.3390/mi11050456}}

@inproceedings{SingleShotOCT,
author = {Molly Subhash Hrebesh and Yuuki Watanabe and Razvan Dabu and Manabu Sato},
title = {{Single-shot full-field OCT based on four quadrature phase-stepped interferometer}},
volume = {6847},
booktitle = {Coherence Domain Optical Methods and Optical Coherence Tomography in Biomedicine XII},
editor = {Joseph A. Izatt and James G. Fujimoto and Valery V. Tuchin},
organization = {International Society for Optics and Photonics},
publisher = {SPIE},
pages = {175 -- 182},
year = {2008},
doi = {10.1117/12.759615},
URL = {https://doi.org/10.1117/12.759615}
}

@ARTICLE{Reza:JSSC,
  author={R. {Fatemi} and A. {Khachaturian} and A. {Hajimiri}},
  journal={IEEE Journal of Solid-State Circuits}, 
  title={A Nonuniform Sparse 2-D Large-FOV Optical Phased Array With a Low-Power PWM Drive}, 
  year={2019},
  volume={54},
  number={5},
  pages={1200-1215},
  doi={10.1109/JSSC.2019.2896767}}

@INPROCEEDINGS{LincolnLabFlash:2016,
  author={Aull, Brian F.},
  booktitle={2016 Conference on Lasers and Electro-Optics (CLEO)}, 
  title={Single-photon-sensitive solid-state image sensors for flash lidar}, 
  year={2016},
  volume={},
  number={},
  pages={1-2},
  doi={}}

@inproceedings{LIDARev:2020,
author = {J. K. Doylend and S. Gupta},
title = {{An overview of silicon photonics for LIDAR}},
volume = {11285},
booktitle = {Silicon Photonics XV},
editor = {Graham T. Reed and Andrew P. Knights},
organization = {International Society for Optics and Photonics},
publisher = {SPIE},
pages = {109 -- 115},
year = {2020},
doi = {},
URL = {https://doi.org/10.1117/12.2544962}
}

@article{rogers2021universal,
  title={A universal 3D imaging sensor on a silicon photonics platform},
  author={Rogers, Christopher and Piggott, Alexander Y and Thomson, David J and Wiser, Robert F and Opris, Ion E and Fortune, Steven A and Compston, Andrew J and Gondarenko, Alexander and Meng, Fanfan and Chen, Xia and others},
  journal={Nature},
  volume={590},
  number={7845},
  pages={256--261},
  year={2021},
  publisher={Nature Publishing Group}
}

@article {Kirmani58,
	author = {Kirmani, Ahmed and Venkatraman, Dheera and Shin, Dongeek and Cola{\c c}o, Andrea and Wong, Franco N. C. and Shapiro, Jeffrey H. and Goyal, Vivek K},
	title = {First-Photon Imaging},
	volume = {343},
	number = {6166},
	pages = {58--61},
	year = {2014},
	doi = {10.1126/science.1246775},
	publisher = {American Association for the Advancement of Science},
	issn = {0036-8075},
	URL = {https://science.sciencemag.org/content/343/6166/58},
	eprint = {https://science.sciencemag.org/content/343/6166/58.full.pdf},
	journal = {Science}
}

@article{Kazovsky:87,
author = {Leonid G. Kazovsky and Lyn Curtis and William C. Young and Nim K. Cheung},
journal = {Appl. Opt.},
number = {3},
pages = {437--439},
publisher = {OSA},
title = {All-fiber 90{\textdegree} optical hybrid for coherent communications},
volume = {26},
month = {Feb},
year = {1987},
url = {http://ao.osa.org/abstract.cfm?URI=ao-26-3-437},
doi = {10.1364/AO.26.000437}}

@article{leeb1983,
  title={Realization of 90-and 180 degree hybrids for optical frequencies},
  author={Leeb, Walter R},
  journal={Archiv Elektronik und Uebertragungstechnik},
  volume={37},
  pages={203--206},
  year={1983}
}

@inproceedings{Reiner:1991,
author = {Reiner B. Garreis},
title = {{90 degree optical hybrid for coherent receivers}},
volume = {1522},
booktitle = {Optical Space Communication II},
editor = {Juergen Franz},
organization = {International Society for Optics and Photonics},
publisher = {SPIE},
pages = {210 -- 219},
year = {1991},
doi = {},
URL = {https://doi.org/10.1117/12.46097}
}

@article{Dong:14120hyb,
author = {Po Dong and Chongjin Xie and Lawrence L. Buhl},
journal = {Opt. Express},
number = {2},
pages = {2119--2125},
publisher = {OSA},
title = {Monolithic polarization diversity coherent receiver based on 120-degree optical hybrids on silicon},
volume = {22},
month = {Jan},
year = {2014},
url = {http://www.opticsexpress.org/abstract.cfm?URI=oe-22-2-2119},
doi = {10.1364/OE.22.002119}}

@article{Jeong:10,
author = {Seok-Hwan Jeong and Ken Morito},
journal = {J. Lightwave Technol.},
number = {9},
pages = {1323--1331},
publisher = {OSA},
title = {Novel Optical 90deg Hybrid Consisting of a Paired Interference Based 2x4 MMI Coupler, a Phase Shifter and a 2x2 MMI Coupler},
volume = {28},
month = {May},
year = {2010},
url = {http://jlt.osa.org/abstract.cfm?URI=jlt-28-9-1323}}

@article{Li:13,
author = {Yanlu Li and Roel Baets},
journal = {Opt. Express},
number = {11},
pages = {13342--13350},
publisher = {OSA},
title = {Homodyne laser Doppler vibrometer on silicon-on-insulator with integrated 90 degree optical hybrids},
volume = {21},
month = {Jun},
year = {2013},
url = {http://www.opticsexpress.org/abstract.cfm?URI=oe-21-11-13342},
doi = {10.1364/OE.21.013342}}

@article{wilkinson:1963,
  title={A method of generating functions of several variables using analog diode logic},
  author={Wilkinson, Robert Haydn},
  journal={IEEE Transactions on Electronic Computers},
  number={2},
  pages={112--129},
  year={1963},
  publisher={IEEE}
}

@article{qing:2015,
  title={Fabrication and characterization of novel high-speed InGaAs/InP uni-traveling-carrier photodetector for high responsivity},
  author={Qing-Tao, Chen and Yong-Qing, Huang and Jia-Rui, Fei and Xiao-Feng, Duan and Kai, Liu and Feng, Liu and Chao, Kang and Jun-Chu, Wang and Wen-Jing, Fang and Xiao-Min, Ren},
  journal={Chinese Physics B},
  volume={24},
  number={10},
  pages={108506},
  year={2015},
  publisher={IOP Publishing}
}

@article{Wang:19,
author = {Youmin Wang and Guangya Zhou and Xiaosheng Zhang and Kyungmok Kwon and Pierre-A. Blanche and Nicholas Triesault and Kyoung-sik Yu and Ming C. Wu},
journal = {Optica},
number = {5},
pages = {557--562},
publisher = {OSA},
title = {2D broadband beamsteering with large-scale MEMS optical phased array},
volume = {6},
month = {May},
year = {2019},
url = {http://www.osapublishing.org/optica/abstract.cfm?URI=optica-6-5-557},
doi = {10.1364/OPTICA.6.000557}}

@article{TUANTRANONT:01,
title = {Optical beam steering using MEMS-controllable microlens array},
journal = {Sensors and Actuators A: Physical},
volume = {91},
number = {3},
pages = {363-372},
year = {2001},
note = {Proceedings of the Technical Digest of the 2000 Solid-State Sensors and Actuators Workshop},
issn = {0924-4247},
doi = {https://doi.org/10.1016/S0924-4247(01)00609-4},
url = {https://www.sciencedirect.com/science/article/pii/S0924424701006094},
author = {Adisorn Tuantranont and V.M. Bright and J. Zhang and W. Zhang and J.A. Neff and Y.C. Lee}}

@article{AflatouniPrj:15,
author = {Firooz Aflatouni and Behrooz Abiri and Angad Rekhi and Ali Hajimiri},
journal = {Opt. Express},
number = {16},
pages = {21012--21022},
publisher = {OSA},
title = {Nanophotonic projection system},
volume = {23},
month = {Aug},
year = {2015},
url = {http://www.opticsexpress.org/abstract.cfm?URI=oe-23-16-21012},
doi = {10.1364/OE.23.021012}}

@Article{Watts:Nature,
author={Sun, Jie
and Timurdogan, Erman
and Yaacobi, Ami
and Hosseini, Ehsan Shah
and Watts, Michael R.},
title={Large-scale nanophotonic phased array},
journal={Nature},
year={2013},
day={09},
publisher={Nature Publishing Group, a division of Macmillan Publishers Limited. All Rights Reserved. SN  -},
volume={493},
pages={195-199},
url={http://dx.doi.org/10.1038/nature11727},
doi={https://doi.org}}

@inproceedings{Wagner:19,
author = {K.H. Wagner and N. Dostart and B. Zhang and M. Brand and D. Feldkhun and M. Popovi\'{c}},
booktitle = {Imaging and Applied Optics 2019 (COSI, IS, MATH, pcAOP)},
journal = {Imaging and Applied Optics 2019 (COSI, IS, MATH, pcAOP)},
pages = {CTh2A.5},
publisher = {Optical Society of America},
title = {SCALABLE: Self-Calibrated Adaptive LIDAR Aperture Beamsteering Light Engine},
year = {2019},
url = {http://www.osapublishing.org/abstract.cfm?URI=COSI-2019-CTh2A.5},
doi = {10.1364/COSI.2019.CTh2A.5},
}

@inproceedings{Miller:18,
author = {Steven A. Miller and Christopher T. Phare and You-Chia Chang and Xingchen Ji and Oscar A. Jimenez Gordillo and Aseema Mohanty and Samantha P. Roberts and Min Chul Shin and Brian Stern and Moshe Zadka and Michal Lipson},
booktitle = {Conference on Lasers and Electro-Optics},
journal = {Conference on Lasers and Electro-Optics},
pages = {JTh5C.2},
publisher = {Optical Society of America},
title = {512-Element Actively Steered Silicon Phased Array for Low-Power LIDAR},
year = {2018},
url = {http://www.osapublishing.org/abstract.cfm?URI=CLEO_QELS-2018-JTh5C.2},
}

@inproceedings{Fatemi:17,
author = {Reza Fatemi and Behrooz Abiri and Ali Hajimiri},
booktitle = {Conference on Lasers and Electro-Optics},
journal = {Conference on Lasers and Electro-Optics},
pages = {JW2A.9},
publisher = {Optical Society of America},
title = {An 8{\texttimes}8 Heterodyne Lens-less OPA Camera},
year = {2017},
url = {http://www.osapublishing.org/abstract.cfm?URI=CLEO_AT-2017-JW2A.9},
doi = {10.1364/CLEO_AT.2017.JW2A.9}}

@INPROCEEDINGS{Chung:17,
  author={Chung, SungWon and Abediasl, Hooman and Hashemi, Hossein},
  booktitle={2017 IEEE International Solid-State Circuits Conference (ISSCC)}, 
  title={15.4 A 1024-element scalable optical phased array in 0.18µm SOI CMOS},
  year={2017},
  volume={},
  number={},
  pages={262-263},
  doi={10.1109/ISSCC.2017.7870361}}

@ARTICLE{Kim:19,  
  author={Kim, Taehwan and Bhargava, Pavan and Poulton, Christopher V. and Notaros, Jelena and Yaacobi, Ami and Timurdogan, Erman and Baiocco, Christopher and Fahrenkopf, Nicholas and Kruger, Seth and Ngai, Tat and Timalsina, Yukta and Watts, Michael R. and Stojanović, Vladimir},
  journal={IEEE Journal of Solid-State Circuits},   
  title={A Single-Chip Optical Phased Array in a Wafer-Scale Silicon Photonics/CMOS 3D-Integration Platform},   
  year={2019},  
  volume={54},  
  number={11},  
  pages={3061-3074},  
  doi={10.1109/JSSC.2019.2934601}}



%

%

\begin{IEEEbiography}[{\includegraphics[width=1in,height=1.25in,clip,keepaspectratio]{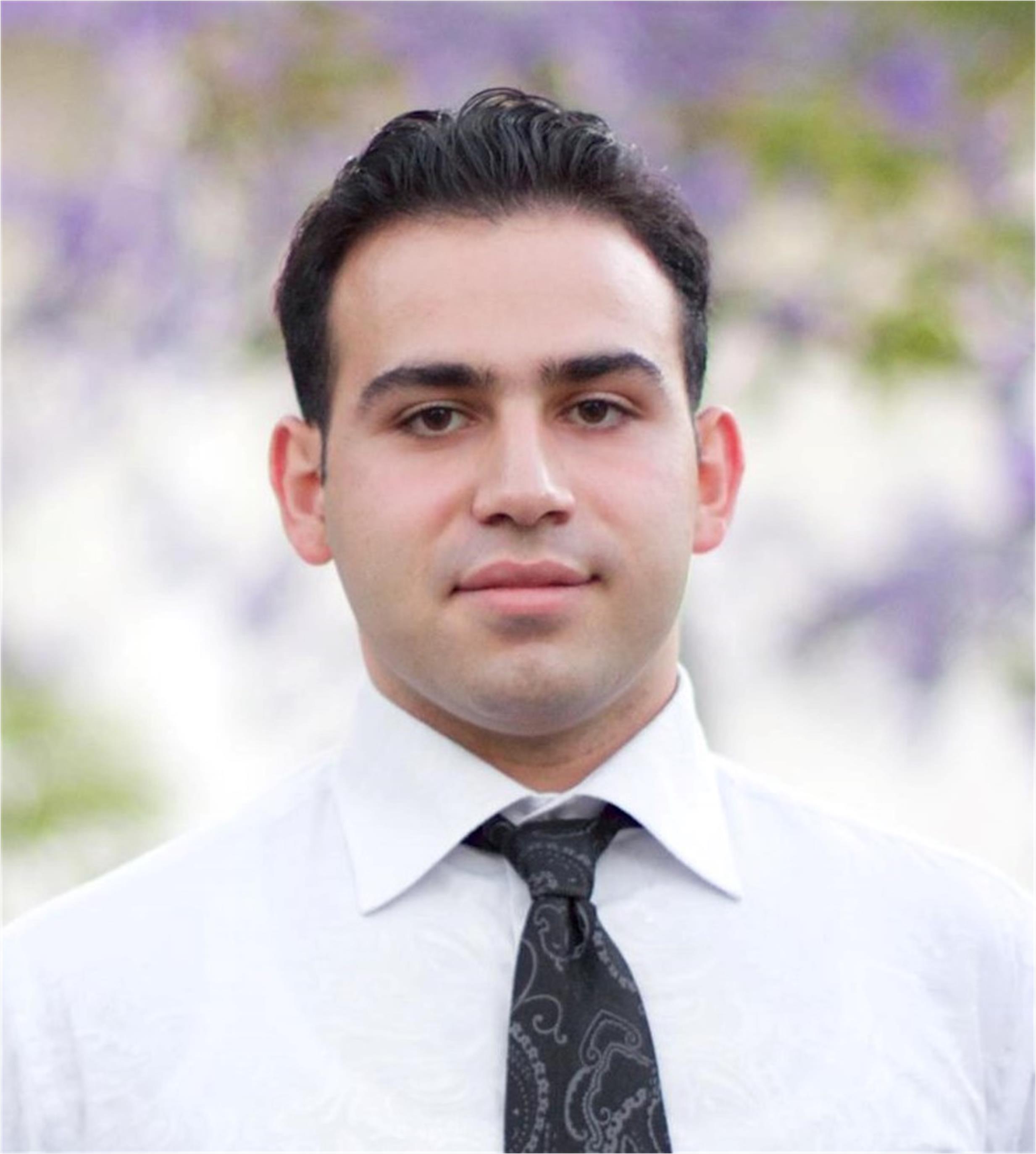}}]{Aroutin Khachaturian}

received the B.S. and M.S., and Ph.D. degrees in electrical engineering from the California Institute of Technology (Caltech) in 2013, 2014, and 2020 respectively. He is currently a post-doctoral scholar in electrical engineering at Caltech. He was a recipient of the Killgore Fellowship and the Analog Devices, Inc. Outstanding Student Designer Award in 2013. His research interests are in the design of integrated electro-optics systems for high-speed optical interconnects and photonic beamforming techniques for communications, imaging, and remote sensing.

\end{IEEEbiography}

\vspace{5mm}

\begin{IEEEbiography}[{\includegraphics[width=1in,height=1.25in,clip,keepaspectratio]{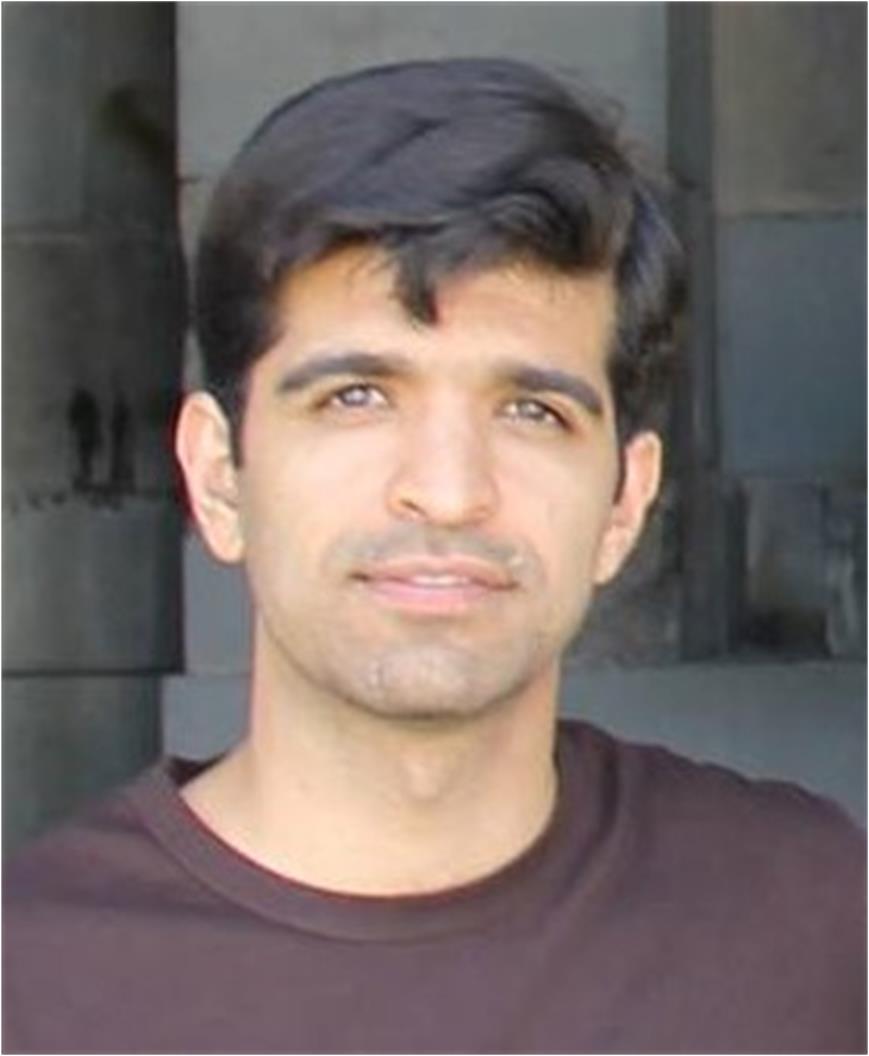}}]{Reza Fatemi}

received the B.S. and M.S. degrees in Electrical Engineering from K. N. Toosi University of Technology, Tehran, Iran in 2011 and Sharif University of Technology (SUT), Tehran, Iran in 2013, respectively. He received his Ph.D. degree in Electrical Engineering at the California Institute of Technology (Caltech) in 2020. He was awarded the Oringer and Kilgore Fellowship in 2014 and the Analog Devices Outstanding Student Designer Award in 2015. His research interests are in silicon photonics and high-speed integrated circuit design.

\end{IEEEbiography}



\begin{IEEEbiography}[{\includegraphics[width=1in,height=1.25in,clip,keepaspectratio]{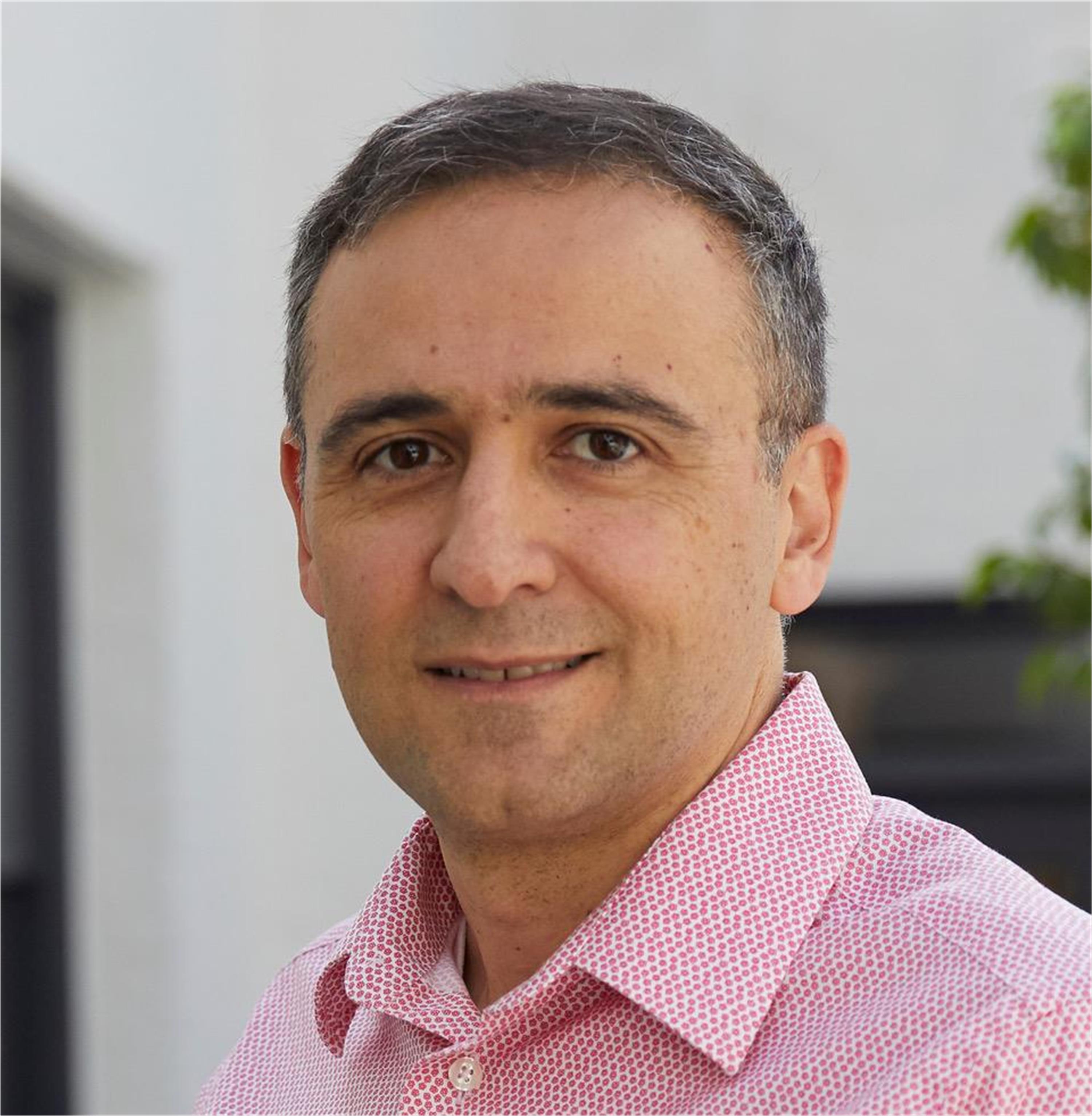}}]{Ali Hajimiri}

(S\textquotesingle 94,M\textquotesingle 98,SM\textquotesingle 09,F\textquotesingle 10) received the B.S. degree in electronics engineering from the Sharif University of Technology, and M.S. and Ph.D. degrees in electrical engineering from Stanford University, Stanford, CA, USA, in 1996 and 1998, respectively. 

He has been with Philips Semiconductors, where he worked on a BiCMOS chipset for GSM and cellular units from 1993 to 1994. In 1995, he was with Sun Microsystems working on the UltraSPARC microprocessors cache RAM design methodology. During the summer of 1997, he was with Lucent Technologies (Bell Labs), Murray Hill, NJ, USA, where he investigated low-phase-noise integrated oscillators. In 1998, he joined the faculty of the California Institute of Technology, Pasadena, where he is the Bren Professor of Electrical Engineering and Medical Engineering and the Director of the Microelectronics Laboratory and Co-Director of Space Solar Power Project. His research interests are high-speed and high-frequency integrated circuits for applications in sensors, photonics, biomedical devices, and communication systems. 

Prof. Hajimiri is a Fellow of National Academy of Inventors (NAI). He was selected to the TR35 top innovator\textquotesingle s list. He is also a Fellow of IEEE and has served as a Distinguished Lecturer of the IEEE Solid-State and Microwave Societies. He is the author of ``Analog: Inexact Science, Vibrant Art'' (2020, Early Draft) a book on fundamental principles of analog circuit design and “The Design of Low Noise Oscillators” (Boston, MA: Springer). He has authored and coauthored more than 250 refereed journal and conference technical articles and has been granted more than 130 U.S. patents with many more pending applications.
He won the Feynman Prize for Excellence in Teaching, Caltech’s most prestigious teaching honor, as well as Caltech\textquotesingle ’s Graduate Students Council Teaching and Mentoring award and the Associated Students of Caltech Undergraduate Excellence in Teaching Award. He was the Gold medal winner of the National Physics Competition and the Bronze Medal winner of the 21st International Physics Olympiad, Groningen, Netherlands. He was recognized as one of the top-10 contributors to ISSCC. He was a co-recipient of the IEEE Journal of Solid-State Circuits Best Paper Award of 2004, the International Solid-State Circuits Conference (ISSCC) Jack Kilby Outstanding Paper Award, a co-recipient of RFIC best paper award, a two-time co-recipient of CICC best paper award, and a three-time winner of the IBM faculty partnership award as well as National Science Foundation CAREER award and Okawa Foundation award. In 2002, he co-founded Axiom Microdevices Inc., whose fully-integrated CMOS PA has shipped around 400,000,000 units, and was acquired by Skyworks Inc. in 2009. He has served on the Technical Program Committee of the International Solid-State Circuits Conference (ISSCC), as an Associate Editor of the IEEE Journal of Solid-State Circuits (JSSC), as an Associate Editor of IEEE Transactions on Circuits and Systems (TCAS): Part-II, a member of the Technical Program Committees of the International Conference on Computer Aided Design (ICCAD), Guest Editor of the IEEE Transactions on Microwave Theory and Techniques, and Guest Editorial Board of Transactions of Institute of Electronics, Information and Communication Engineers of Japan (IEICE). 

\end{IEEEbiography}




\end{document}